%% file: main.tex
\documentclass[a4paper,conference]{IEEEtran}

\usepackage{cite}      

\usepackage{graphicx}  

\graphicspath{{./figures/}}

\usepackage{subfigure} 

\usepackage{url}       

\usepackage{amsmath}   
\usepackage{fancyhdr}
\usepackage[pdftex,hypertexnames=false]{hyperref}

\newcommand{\swift}{\textsc{swift}}

\begin{document}

\title{\swift{}: Maintaining weak-scalability with a dynamic range of $10^4$ in time-step size to harness extreme adaptivity}

\author{\IEEEauthorblockN{Josh Borrow, Richard G. Bower, Peter W. Draper}
\IEEEauthorblockA{Institute for Computational Cosmology\\
Durham University\\
Durham, United Kingdom\\
joshua.borrow@durham.ac.uk}
\and
\IEEEauthorblockN{Pedro Gonnet}
\IEEEauthorblockA{Google Switzerland GmbH\\
Brandschenkestrasse 110\\
8002 Zurich, Switzerland}
\and
\IEEEauthorblockN{Matthieu Schaller}
\IEEEauthorblockA{Leiden Observatory\\
Leiden University\\
Leiden, Netherlands}}


\maketitle

\begin{abstract}
	\input{abstract}
\end{abstract}

\section{Introduction}
\label{sec:intro}
\input{introduction}

\section{The Necessity of a scheme using multiple time-steps}
\label{sec:problem}
\input{problem}

\section{The Cosmological Simulation code \swift{}}
\label{sec:swift}
\input{swift}

\section{Domain Decomposition}
\label{sec:domain}
\input{domain}

\section{Weak-Scaling and Performance Results}
\label{sec:result}
\input{results}

\section{Conclusions}
\label{sec:conc}
\input{conclusion}
\section*{Acknowledgements}
\label{sec:ack}
\input{acknowledgements}

\bibliographystyle{IEEEtran}
\bibliography{bibliography}

\end{document}

%% file: abstract.tex
Cosmological simulations require the use of a multiple time-stepping scheme. Without such a scheme, cosmological simulations would be impossible due to their high level of dynamic range; over eleven orders of magnitude in density. Such a large dynamic range leads to a range of over four orders of magnitude in time-step, which presents a significant load-balancing challenge. In this work, the extreme adaptivity that cosmological simulations present is tackled in three main ways through the use of the code \swift{}. First, an adaptive mesh is used to ensure that only the relevant particles are interacted in a given time-step. Second, task-based parallelism is used to ensure efficient load-balancing within a single node, using pthreads and SIMD vectorisation. Finally, a domain decomposition strategy is presented, using the graph domain decomposition library METIS, that bisects the \emph{work} that must be performed by the simulation between nodes using MPI. These three strategies are shown to give \swift{} near-perfect weak-scaling characteristics, only losing 25\% performance when scaling from 1 to 4096 cores on a representative problem, whilst being more than 30x faster than the de-facto standard Gadget-2 code.

%% file: introduction.tex
Smoothed Particle Hydrodynamics (SPH) is a numerical method now widely used in many scientific fields. This particle-based method, unlike grid-based methods, has adaptivity `built in'; higher-density regions are automatically represented using a higher particle density than lower-density regions. In a grid-based method, the grid must be adaptively refined to ensure higher resolution in these regions. The adaptive nature of SPH also means that this adaptivity in density also provides appropriate adaptivity in the length-scales that are resolved. This leads naturally to the idea of an adaptive \emph{local} time-step which correlates with the density of the region (via the CFL condition). This ability is often ignored outside of astrophysics simulations where single \emph{global} time-steps are usually used for most science cases.

Cosmological simulations (e.g. \cite{Schaye2015, Hopkins2017,Pillepich2018}) harbour a very large dynamic range, hence requiring high adaptivity in the length-scales resolved. The inclusion of gravitational forces allows perturbations in the initially (nearly) uniform fluid to grow over time, forming galaxies. These galaxies, the regions of interest, are many orders of magnitude more dense than the voids that they leave behind. Whilst the resulting high level of adaptivity takes advantage of a unique feature of unstructured methods such as SPH, several other problems emerge. Even though strategies have been proposed to achieve good scaling in cases with nearly uniform particle distributions (compared to cosmological cases) and global time-steps (e.g. \cite{Oger2016}), maintaining strong- and weak-scalability with such adaptive problems is notoriously difficult.

The focus of this work is to consider the problems that temporal adaptivity, in the form of local time-stepping, poses in detail, and suggest solutions and best practices that are implemented in the cosmological simulation code \swift{}. The structure of the paper is as follows: in \S \ref{sec:problem} the problem posed by time-stepping in cosmological simulations is stated in more detail; in \S \ref{sec:swift} the cosmological simulation code \swift{} is described; in \S \ref{sec:domain} different domain decomposition strategies are considered and compared; and in \S \ref{sec:result} the resulting weak-scaling properties of \swift{} are presented. 

%% file: problem.tex
\begin{figure*}
    \centering
    \includegraphics[width=\textwidth, trim=0cm 0.5cm 0cm 0.6cm]{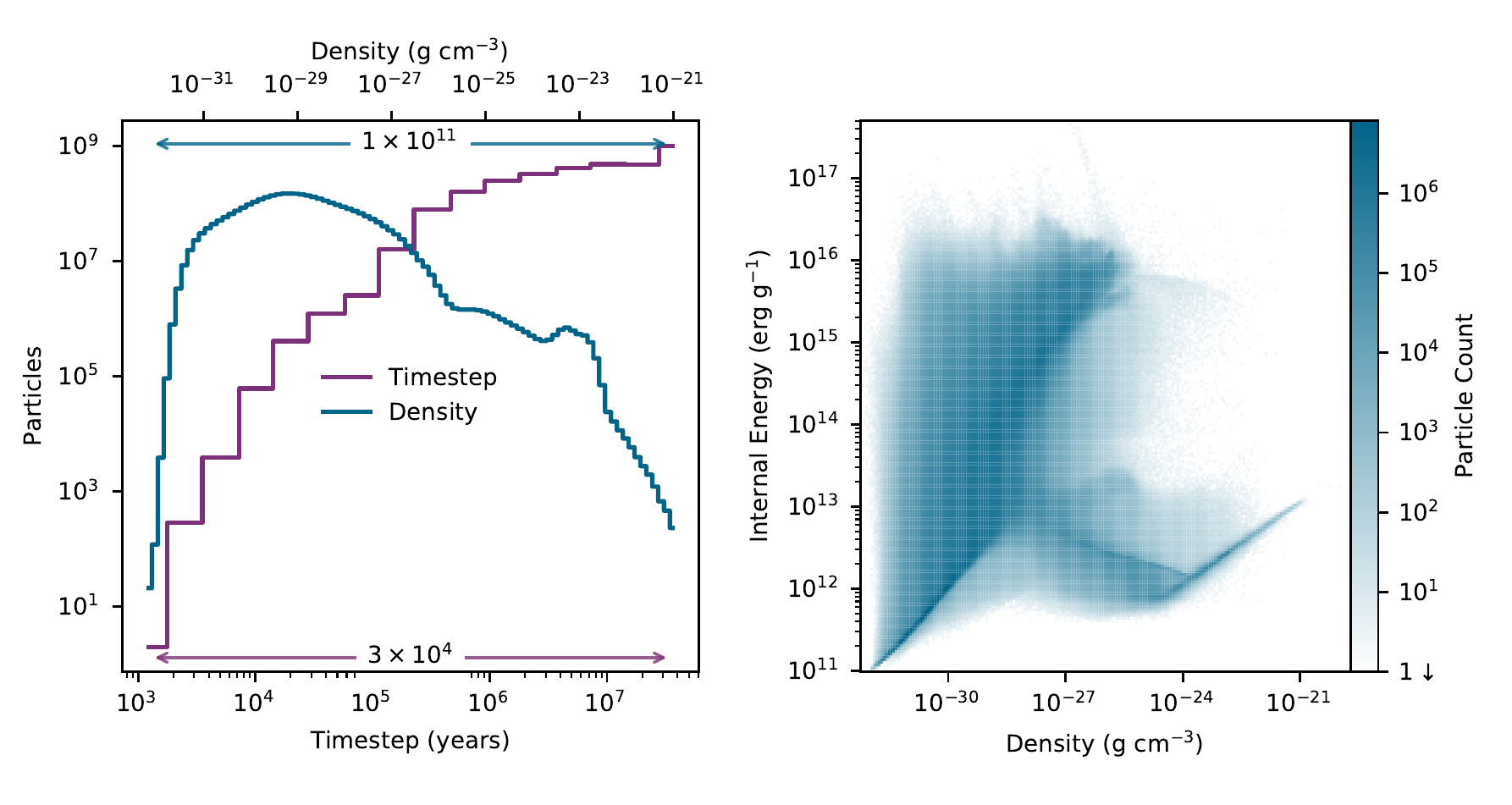}
    \caption{\textit{(left)} Histogram of time-steps and densities in the flagship run from the EAGLE simulation suite \cite{Schaye2015} at redshift $z=0.1$ (near the end of the simulation when bound structures have formed, see http://icc.dur.ac.uk/Eagle for more information on the EAGLE project). Note the logarithmic axis; this shows the large range of time-step sizes that must be captured for an efficient calculation. Also note the low number of particles on those small time-steps; this is not a cumulative histogram. \textit{(right)} A Density-Internal Energy diagram showing the wide variation in particle temperatures and densities in our simulated universe that require appropriate time stepping; this is captured using the CFL condition. Note that the bottom-right diagonal cut-off is imposed by hand to prevent spurious fragmentation of gas clouds inside galaxies (see \cite{Schaye2015} for the physical motivation).\vspace{-0.4cm}}
    \label{fig:problem:dynamic}
\end{figure*}

It is worth pausing for a moment and considering more carefully the specific numerical challenges that cosmological simulations pose. Cosmological simulations are a case of extreme adaptivity. For example, in the EAGLE simulation (see \cite{Schaye2015, Crain2015, Schaller2015}), when the galaxies are fully developed, there is a dynamic range of $10^{11}$ in gas density, and $10^6$ in internal energy. An appropriate time-step size, for a standard SPH scheme (see \cite{Price2012}), is provided by the Courant-Friedrichs-Lewy (CFL) condition \cite{Courant1928},
\begin{align}
    \Delta t_{i} = \mathcal{C}
                   \frac{h_i}
                   {{\rm{max}}_j (c_i + c_j + 
                   {\rm{max}}\{0, -3 \mathbf{r}_{ij} \cdot \mathbf{v}_{ij}/r_{ij}\})} ,
    \label{eqn:problem:cfl}
\end{align}
where $\mathcal{C}$ is the CFL constant, usually set to $\approx 0.2$, $h_i$ is the smoothing length of particle $i$, $\mathbf{v}_{ij} = \mathbf{v}_i - \mathbf{v}_j$, $\mathbf{r}_{ij} = \mathbf{r}_i - \mathbf{r}_j$, $c_i$, and $c_j$ are the fluid velocity between, the distance between, and the sound speeds of the particles $i$ and $j$ respectively. Here, $j$ refers to all neighbours of the particle $i$. This gives an effective scaling of the time-step size with density ($\rho$) and internal energy ($u$) as $\Delta t_{i} \propto u_i^{-1/2}\rho_i^{-1/3}$ for a poly-tropic equation of state, such that hotter and denser particles get smaller time-steps. Applied to the EAGLE simulation, this gives a dynamic range of over $10^4$ in time-step size (see Fig. \ref{fig:problem:dynamic}). It is worth noting that the CFL condition is not the only condition on the time-step of a particle in a cosmological simulation: self-gravity, cooling, and other sub-grid processes all enter the time-step calculation but mostly affect the already dense regions. Another feature shown in Fig. \ref{fig:problem:dynamic} is that there are many orders of magnitude fewer particles on the small time-steps (that must be updated frequently) than those on long time-steps. The typical strategy is to evolve \emph{all} particles using the lowest time-step in the simulation (also referred to as a \emph{global} time-step). However, given the depth of the time-step hierarchy shown here, this would increase the runtime of a given problem by many orders of magnitude. Hence, without a scheme allowing for multiple time-steps (also often referred to as \emph{local} time-steps), cosmological simulations would be infeasible.


\subsection{Efficient Implementation of Multiple Time-steps}

In a typical velocity-Verlet kick-drift-kick time-integration algorithm \cite{Verlet1967, Swope1982}, the multi-time-stepping scheme is realised by assigning each particle an individual time-step, and only calculating forces and accelerations (the `kick' step) for each particle once the simulation time proceeds to the next update time of that particle. This method, widely used for pure $N$-body problems, can be applied to any system where the Hamiltonian can be split into two terms acting on the positions (the `drift') and velocities (the `kick') separately, as is the case for SPH. This allows for a significant reduction in computation for a cosmological problem, but is more tricky to implement than a global time-stepping scheme. For many years in cosmology, since the introduction of the TreeSPH code \cite{Hernquist1989}, time-step-binning has been used. Particles are assigned to a corresponding time-bin ($n$) such that their time-step,
\begin{align}
    \Delta t_i \geq 2^n \cdot t_{\rm min},
    \label{eqn:swift:binning}
\end{align}
is discretized with respect to the absolute minimal time-step in the simulation $t_{\rm min} = 2^{-N_{\rm bin}} t_{\rm run}$, with $t_{\rm run}$ the simulation time for the whole run, $t_{\rm end} - t_{\rm begin}$, and $N_{\rm bin}$ the total number of time-bins required. At each time-step in the code all particles with time-bin $n<m$, with $m$ the current maximal active time-bin, are kicked. A particle can be drifted as many times between now and the next time it is active, as long as all drifts ensure that the particle `experiences' $\Delta t_i$ between now and then (e.g. drift with $\Delta t_i / 2^{j}$, $j$ times). The code then needs only to iterate through all occupied time-bins until all particles have been updated to synchronise the simulation.

One possible remedy to the high dynamic range in time-step is simply to `drift' (i.e. update the position of) every particle each time-step (a relatively cheap operation) whilst only `kicking' the active particles (a more expensive operation since it involves looping over neighbours). Since all particles have been drifted to the current time, all the neighbours of the particles being kicked are, by construction, at their current position, ensuring that the loops over neighbours use the current state of the neighbours. This popular hybrid strategy, used by many codes (e.g. Gadget-2 \cite{Springel2005}, SEREN \cite{Hubber2011}, Gasoline2 \cite{Wadsley2017GASOLINE2:Code}, Phantom \cite{Price2017}), is quite effective as it drastically reduces the number of loops over neighbours required since only loops for active particles are computed. It also naturally leads to a high number of operations per second (i.e. an apparent effective use of the system) and to good scalability since the drift operation is very simple to parallelize. In some cases, the `drift' operation becomes the main hotspot of the calculation. However, few of those operations are \emph{useful} operations, as most of the drifted particles will \emph{not} be neighbours of any active particles. This is especially true in cases where the time-step hierarchy becomes very deep, as in cosmological simulations or simulations of the formation of the collapse of a gas cloud turning into stars.

This demonstrates that for such adaptive simulations it is easy to generate spurious `FLOPS', making such a metric irrelevant to any discussion around this class of problems. We therefore argue that the only relevant metric in this space is the time-to-solution for representative adaptive problems.

\begin{figure}
    \centering
    \includegraphics[width=\columnwidth, trim=0cm 0.5cm 0cm 0.5cm]{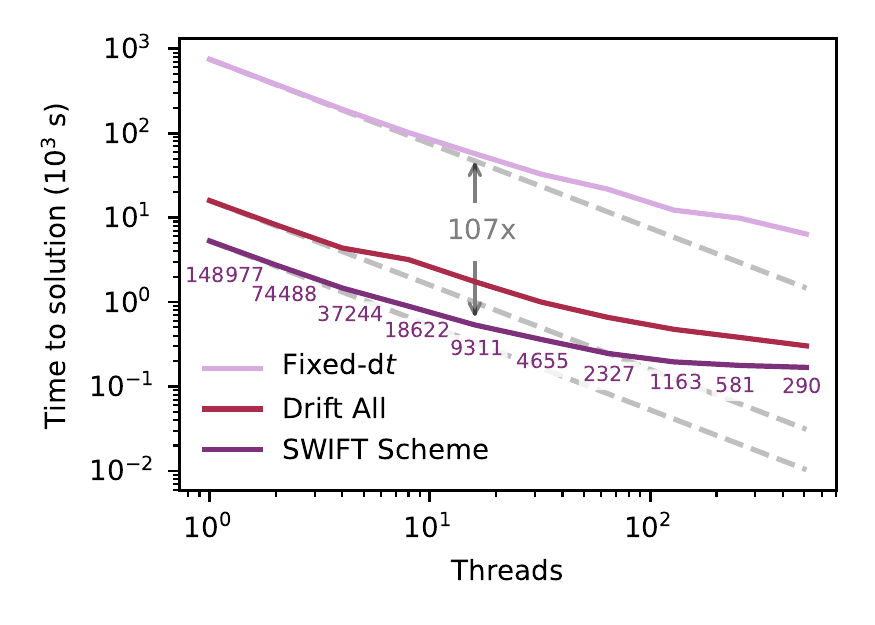}
    \caption{A comparison of the time-to-solution when running with a global time-step, a scheme that drifts all particles each time-step (but kicks only the active ones), and the optimal strategy, which is to only drift relevant particles. All three results were obtained with the \swift{} code in different running modes. Note the over 100x \emph{gain} in time-to-solution on this representative problem (a $53\times10^6$ particle box from the EAGLE project at redshift $z=0.1$, Fig. \ref{fig:problem:dynamic}), when updating and drifting only the relevant particles. The grey dashed lines in the background indicate perfect scaling. The purple labels indicate the average number of \emph{active} (kicked) particles per time-step per core; note how this approaches merely hundreds when attempting to strong-scale such a problem to more than 100 cores, thus severely limiting strong-scaling. 
    \label{fig:problem:strong}}
\end{figure}

An efficient code in the context of a deep time-step hierarchy hence ought to reduce or completely eliminate the drift operations for particles that are not directly involved in the calculation of accelerations (i.e. direct neighbours of an active particle being kicked). Unfortunately, a direct application of Amdahl's law \cite{Amdahl1967} in this scenario implies that such a code will be less scalable. However, since the time spent drifting particles used to dominate the total, the time-to-solution will improve. This is illustrated in Fig. \ref{fig:problem:strong}, where different strategies for the same problem are compared in terms of time-to-solution. The scheme updating and drifting only the relevant particles (labelled as `SWIFT Scheme') is more than 100x faster than the standard strategy despite displaying worse scaling properties than the more naive scheme.

The challenge for the developers then becomes how to load-balance operations that may only need to update (now both for kicks \emph{and} drifts) fewer particles than there are compute cores on the system. This is a non-trivial problem as is exemplified by the very-low average number of updates per core per time-step shown by the labels on Fig. \ref{fig:problem:strong}. The remainder of this paper discusses how this is performed in \swift{}, as well as how the code avoids drifting all the particles at every time-step.

%% file: swift.tex
\begin{figure}
    \centering
    \includegraphics[width=\columnwidth, trim=0cm 0.5cm 0cm 0.5cm]{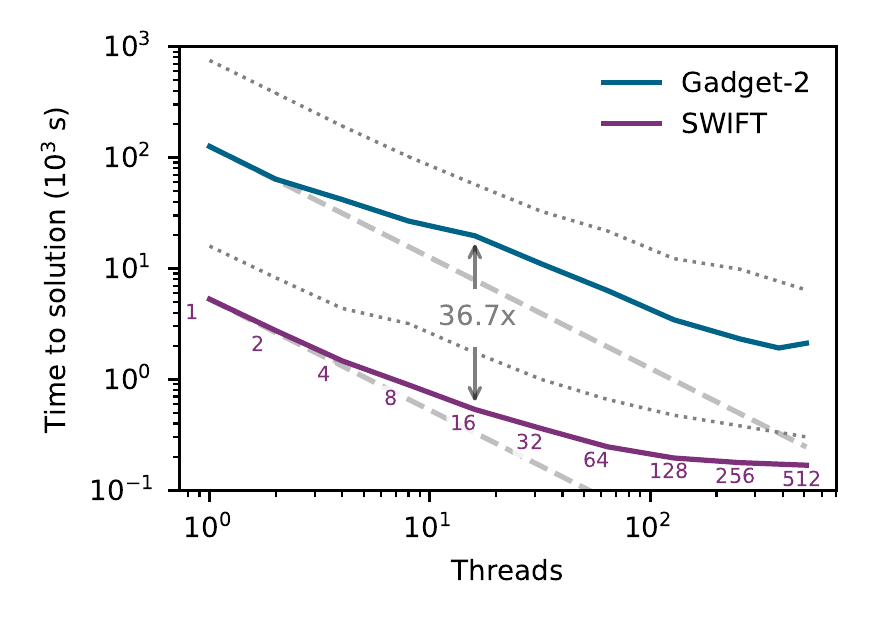}
    \caption{The run time behaviour of \swift{} when strong-scaling. This plot shows an adaptive problem with $6\times10^7$ particles, with an effective `speed-up' of around 35 times when using \swift{} as a drop-in replacement for the de-facto standard Gadget-2. The particles are taken from the EAGLE simulation at redshift $z=0.1$. \swift{} does not scale as well as Gadget-2 at intermediate core counts due to the fact that it does not drift all particles each time-step. It does, however, have a considerably faster time-to-solution than Gadget-2 due to the lack of this spurious work, among other characteristics explored below. The purple text indicates the thread counts of each individual run, with the grey dotted lines showing the results from Fig. \ref{fig:problem:strong}.\vspace{-0.4cm}}
    \label{fig:swift:strong}
\end{figure}

\swift{} \cite{Schaller2016} is a hybrid MPI \& threads C99 code that implements several SPH and particle-based hydrodynamical schemes, a Fast-Multipole-Method (FMM) N-body gravity scheme \cite{Cheng1999, Dehnen2014}, and several sub-grid galaxy formation models, notably the EAGLE 	\cite{Schaye2015}, GEAR \cite{Revaz2011}, and GRACKLE \cite{Smith2017} models. \swift{}\footnote{ More information, documentation, examples, and an automated test-suite is available on our project web pages \url{https://www.swiftsim.com/}} is completely open source and is in open development. The code is designed with next-generation systems in mind, and as such includes several significant improvements over previous-generation codes, which are detailed below.

\swift{} is a hybrid code exploiting all three levels of parallelism available on modern CPU-based architectures: Asynchronous MPI communications are used between cluster nodes, POSIX pthreads are used within individual node \cite{Gonnet2015}, and SIMD instructions are employed to exploit core-level parallelism. More information on the (AVX, AVX2, AVX-512) vectorized routines present in \swift{}, used for neighbour finding and other common operations, can be found in \cite{Willis2017}. 

The run time behaviour of \swift{} when strong-scaling with respect to the leading code in the field, Gadget-2 \cite{Springel2005}, is shown using a representative cosmological problem in Fig. \ref{fig:swift:strong}.

\subsection{Cell Structure}

\begin{figure}
    \centering
    \includegraphics[width=\columnwidth]{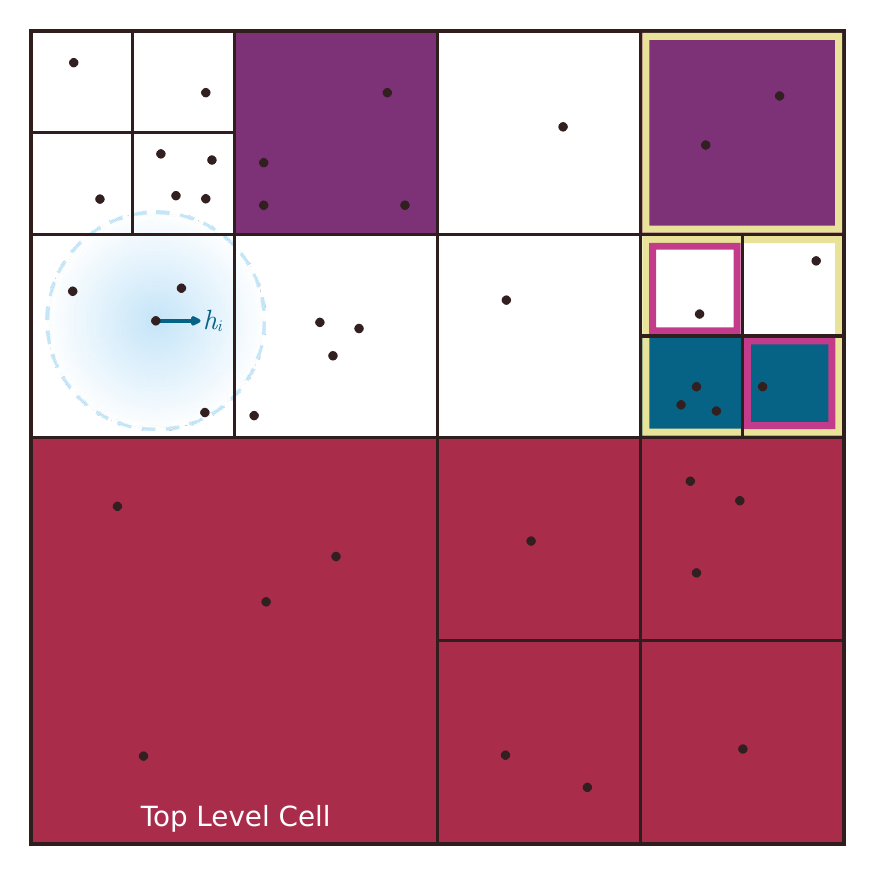}
    \caption{The AMR grid super-structure that is imposed in \swift{}. Interactions need to be computed between all neighbouring cells, and `self' interactions need to be computed within a single cell. Here, several examples are shown. Cells interact on the smallest similar scale possible. Due to the geometrical condition used to build the mesh, it is ensured that the two purple cells cannot interact even though their top-level cells touch (a typical smoothing length and kernel are shown in light blue). On the largest scale, the red (top-level) cells are shown to interact; note that they must interact on the \emph{largest common scale}, and hence all four sub-cells on the right must be included in the calculation. On the smallest scale, the blue interacting cells are shown. Outlined in yellow and purple are two \emph{conflicts} with the `pair' calculation between the two blue cells. These tasks require the same data, so cannot be computed at the same time. However, they need not happen in any particular order. \vspace{-0.4cm}}
    \label{fig:swift:cells}
\end{figure}

\swift{} uses an adaptive cell structure (see Fig. \ref{fig:swift:cells}) for efficient neighbour finding through a pseudo-Verlet list \cite{Gonnet2013} specifically tailored for SPH with a large dynamic range \cite{spheric2013, Gonnet2015}. The simulation is initialised with a fixed number of top-level cells that are adaptively refined based on their particle contents until a given cell contains less than 400 particles. Such a construction ensures that a particle in a cell need only interact with other particles in that cell, or particles in cells that directly neighbour it. It also ensures that the content of a pair of cells neatly fits into the low-level caches of the compute cores. 

At every time-step, a list of the cells containing active particles is constructed. All interactions between pairs of cells who have at least one member on the list are computed. These pairs, alongside the individual cells containing active particle, are distributed over the various threads and the interactions between the particles within a given cell-pair are computed in parallel using vector instructions \cite{Willis2017}. The problem then becomes the load-balancing of this cell-pairs between threads within nodes and across MPI ranks, especially in the cases where very few particles are active.


\subsection{Task-based Parallelism}

\begin{figure*}[ht]
    \centering
    \includegraphics[width=\textwidth, trim=0cm 0cm 0cm 0.5cm]{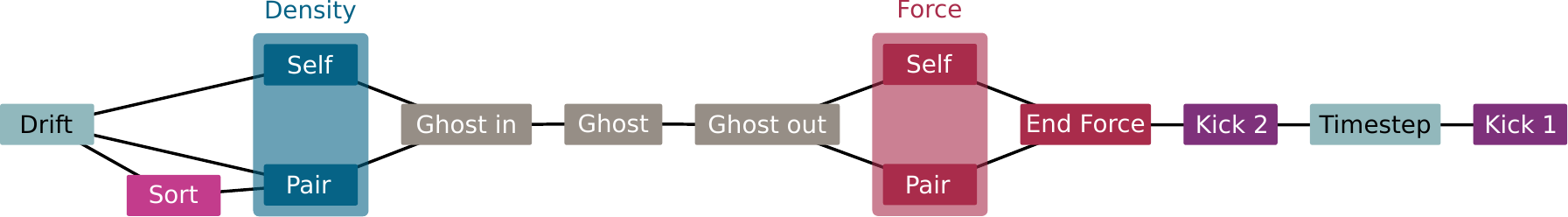}
    \caption{Task dependency-plot for SPH in \swift{}, with execution flowing right. Note that the black lines represent dependencies; conflicts (that would take place in the shaded regions for Density and Force) are not shown. These tasks also interact with tasks from other portions of the code, such as the gravity solver, however these are omitted for clarity. The `ghost' task, among other things, corrects the smoothing length for each particle if required. The `sort' task sorts particles as part of the neighbour-finding scheme (see \cite{Gonnet2015}). For more advanced schemes, such as SPH-ALE \cite{Vila1999,Hopkins2015} or SPH schemes using more loops over neighbours (e.g. \cite{Schaller2015}), an extra loop in the form of pair and self tasks are added.\vspace{-0.4cm}}
    \label{fig:swift:tasks}
\end{figure*}

\swift{} implements task-based parallelism using a modified version of the QuickSched\footnote{\url{https://gitlab.cosma.dur.ac.uk/swift/quicksched}} library \cite{Gonnet2016}. QuickSched deviates from most available tasking libraries through the inclusion of task \emph{conflicts}, not just dependencies. A conflict is generated whenever two tasks require the same data, but may be executed in any order. This concept of conflicts interacts with the cell structure when considering interactions between a cell and its neighbours; these interactions may proceed in any order but should not be executed by two threads at the same time in order to avoid overwriting the same particle acceleration, for example.  Enforcing a dependency for such a situation, which other libraries require the user to do, imposes a spurious ordering to the tasks and may reduce efficiency and create a more complex task-graph. The tasking library also enables easy integration with other types of physics, such as gravity, that are enforced as a different type of task. A plot of the part of the task-graph for SPH in \swift{} is shown in Fig. \ref{fig:swift:tasks}.

\subsection{Asynchronous Communications}

In the scenario where a neighbouring cell `lives' on a different node, communications over MPI must be initiated. However, instead of the traditional `halo model' where data is exchanged before the start of a time-step, implying a synchronisation point, \swift{} handles everything through the tasking system following \cite{Chalk2017}. All nodes contain the cell graph of themselves and their adjacent nodes such that they will be able to generate the correct send and receive tasks. These communications are simply modelled as another task within the system. The asynchronous communications ensure that the compute units do not sit idle during the time it takes for the data to come over the network; purely local work, such as the contribution of purely local particles to the density in the domain, is scheduled whilst the information is arriving. The tasks that depend on data from foreign cells are scheduled after the data has arrived and can start work straight away. If there is not enough local work for each node to perform, e.g. there is a poor domain decomposition where a dense cluster is split down the middle, then the system will be forced to wait.

\subsection{Time-stepping}

The cell structure in \swift{} provides a convenient way to only drift relevant particles. At the start of a time-step (i.e. for a given time-bin and all time-bins smaller than that), the cell-tree is walked in parallel (using threads) and all the tasks linked to cells that contain active particles (and their neighbours) are marked as active. These cells are guaranteed (see Fig. \ref{fig:swift:cells}) to be the only ones that contain particles that are neighbours to the active particles; as such, they contain all of the particles that require drifting during this time-step. When there are orders of magnitude more inactive particles than active particles, this provides a significant gain over the naive implementation where all particles are drifted every time-step (Fig. \ref{fig:problem:strong}).

%% file: domain.tex
Using the scheme described above, the overall amount of work that needs to be completed by a given node is drastically reduced. This presents a problem for large simulations that are spread over many nodes. On the one hand. it is still important to ensure that the longest time-steps, where \emph{all} particles are drifted and then kicked, are well load-balanced, as these will take up the majority of the computation time. But, on the other hand, to ensure efficient computation of the particles on small time-steps, the system should be balanced such that these particles require as little communication as possible; with so little work, the network can end up adding significant latency.

Getting a `good' domain decomposition is key to weak-scaling efficiently on highly adaptive problems. In problems with low dynamic range it may suffice to use a simple grid decomposition, where particles are spread across nodes spatially. In cosmological codes, such as RAMSES \cite{Teyssier2001} (AMR grid code) and Gadget-2 \cite{Springel2005}, it is also common to use a spatial domain decomposition with a grid-like structure. In practice, Peano-Hilbert space-filling curves transform the 3D problem into a single dimension. The 1D string can then be cut to ensure that a similar number of particles remains on each node, whilst reducing the surface of each domain (a proxy for communication cost). This does not, however, guarantee efficient computation, as the balance of \emph{work} across the nodes may not be well distributed. Typically these schemes also include some effort to add a basic weighting to move work, but the nature of the space-filling curve method only ensures that the \emph{mean} work is well distributed; it says nothing about the tail of nodes that have a larger amount of work than average that block execution until they are finished. These Peano-Hilbert schemes also usually reduce the amount of communication in all steps. In \swift{} the amount of communication is almost inconsequential as it is performed asynchronously\footnote{Unless, of course, a (bad) domain decomposition algorithm places a region dominated by small time-steps right next to a domain-edge.}. The only time when communication balancing is important is during the smallest steps.
	
    
We also note that thanks to the hybrid MPI and threads parallelization that \swift{} uses, the typical sizes of domains are an order of magnitude larger than a pure MPI implementation (one domain per node vs. one per core). This means that an order of magnitude fewer domains are required, and makes the task of appropriately portioning the simulation much easier.


\subsection{Domain Decomposition Strategies}

The simplest domain decomposition that is possible is probably to construct a grid of $N^3$ cells (each of the same size), with $M$ is the number of domains required. The same number of cells is associated with each domain, meaning each domain has $N^3/M$ cells with a regular pattern. This grid is then overlaid spatially on the particle positions, and particles are assigned to a node that corresponds to the cell that they lie within. This approach has a number of non-ideal consequences; there may be many more particles on one node than another, for instance, which is highly memory-inefficient. In the following discussion, this will be referred to as the `grid' domain decomposition strategy.

As a thought experiment, the ideal way to produce domains in a cosmological context would be to identify regions, where time-steps will be smallest (typically at the centre of galaxies), and grow domains around them with a watershed-like algorithm. This has two main drawbacks: identifying galaxies is both conceptually and computationally difficult, and this strategy is highly specific to galaxy formation problems. This strategy also runs into problems when two such galaxies merge as a domain-edge would then directly cross the newly formed object.

\begin{table}
    \centering
	\caption{The main cell/edge weighting schemes available in \swift{}}
    \begin{tabular}{|l | l | l | p{2.5cm}|}
    \hline
    \textbf{Name} & \textbf{Cells} & \textbf{Edges} & \textbf{Relevance} \\ \hline
    none/none     & No weighting   & No weighting   & Problems where no particles should be moved and any
	                                              decomposition is adequate. \\ \hline
    costs/costs   & Task costs    & Task costs    & Problems where a balance between communication and computation should be equally considered. \\ \hline
    none/costs    & No weighting   & Task costs    & Problems where communication must be reduced at all
	                                              costs. \\ \hline
    \textbf{costs/time}    & Task costs    & Communication  & Problems where communication must be minimised for
	                                              steps involving few particles, but the problem \emph{must}
						      be well load-balanced for large steps. \\
    \hline
    \end{tabular}
    \label{tab:domain:types}
\end{table}

Therefore a compromise must be reached; a method that is computationally efficient, generic, and that produces a reasonable (not necessarily the \emph{best}) domain decomposition is required. In \swift{} the solution is provided by decomposing the top-level cell graph. Cells are represented by graph nodes assigned a cost dependent on the number and types of tasks associated with that cell (and thus the amount of computation), and dependencies and conflicts between cells are modelled as graph hyper-edges with weights determined from the time of the next particle updates (which is a proxy for activity and hence communication likelyhood).

%
%

\begin{figure}
    \centering
    \includegraphics[width=\columnwidth]{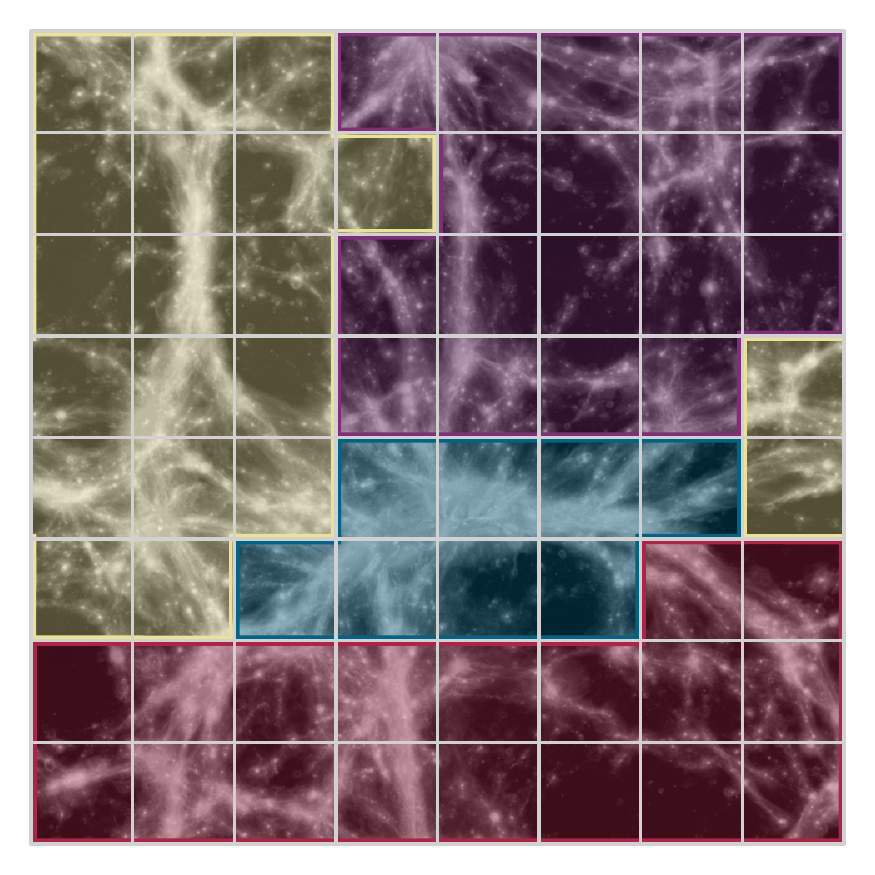}
    \caption{A sketch of an example domain decomposition with METIS \cite{METIS}, using the `costs/time' strategy. A sketch is used as domain decompositions are difficult to visualise in 2D, due to their complex 3D shapes. In the background, a visualisation of the EAGLE simulation is shown at redshift $z=0$ (today). Galaxy clusters, such as the one in the blue domain, lead to naturally smaller domains due to their higher density of particles (and hence increased work). Cuts between domains tend to end up in void-like regions, with very low density. Note the different sized domains that can contain different numbers of particles and be of arbitrary shapes. The grey grid shows the top-level cells. Domains can also span over the edges of the periodic box.\vspace{-0.4cm}}
    \label{fig:domain:illustration}
\end{figure}

This strategy transforms the domain decomposition problem into a standard graph partitioning problem, which can be solved by many software packages. Here, the METIS library \cite{METIS} is chosen as it was the easiest to integrate with the existing code base. The node costs and edge weights are passed to METIS, which then returns a solution for a reasonable partitioning of the \emph{work}. Note that in this system, there is no explicit effort made to balance memory on each node; thankfully work is at least roughly proportional to the number of particles.

In some cases, users may not care about the cost of communications. In \swift{}, thanks to the asynchronous communication during tasks, only very short updates are communication-bound; in a code where there is little range in time-step these should be ignored. For that purpose, and others, several weighting modes are provided. These are described in Table \ref{tab:domain:types}. In the remainder of the text, only the `costs/time' strategy is considered; this is the most relevant (both computationally and physically, see Fig. \ref{fig:domain:illustration}) for a cosmological problem.

\subsection{Comparison of Strategies}

\begin{figure*}
    \centering
    \includegraphics[width=\columnwidth, trim=0cm 0.5cm 0cm 0.6cm]{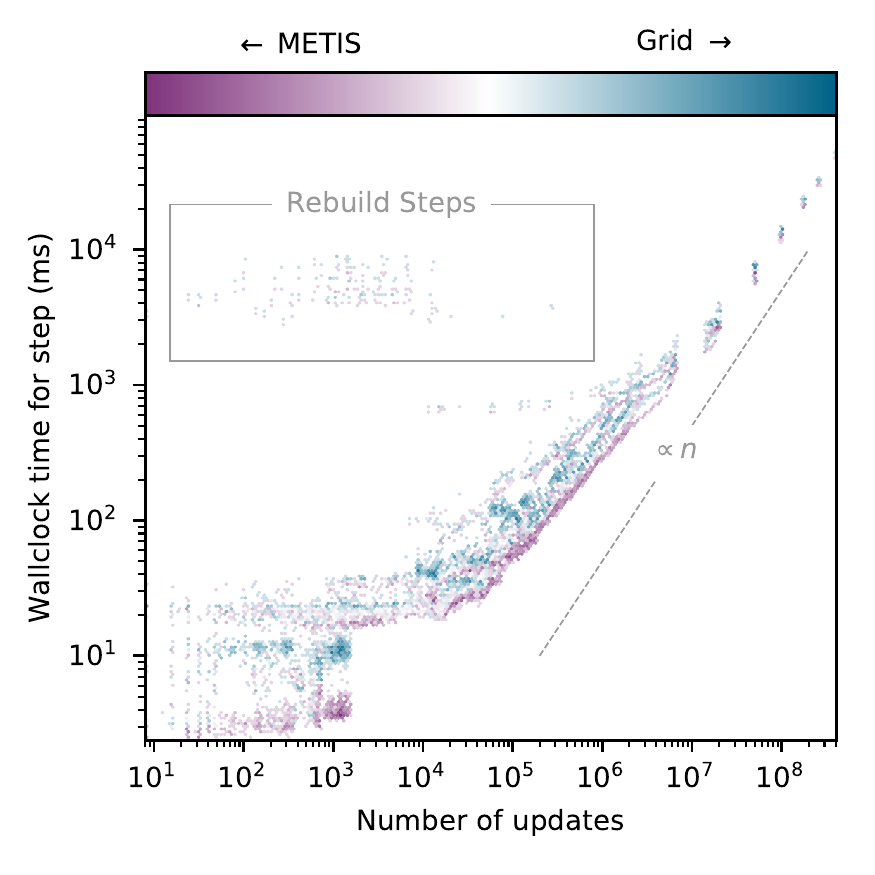}
    \includegraphics[width=\columnwidth, trim=0cm 0.5cm 0cm 0.6cm]{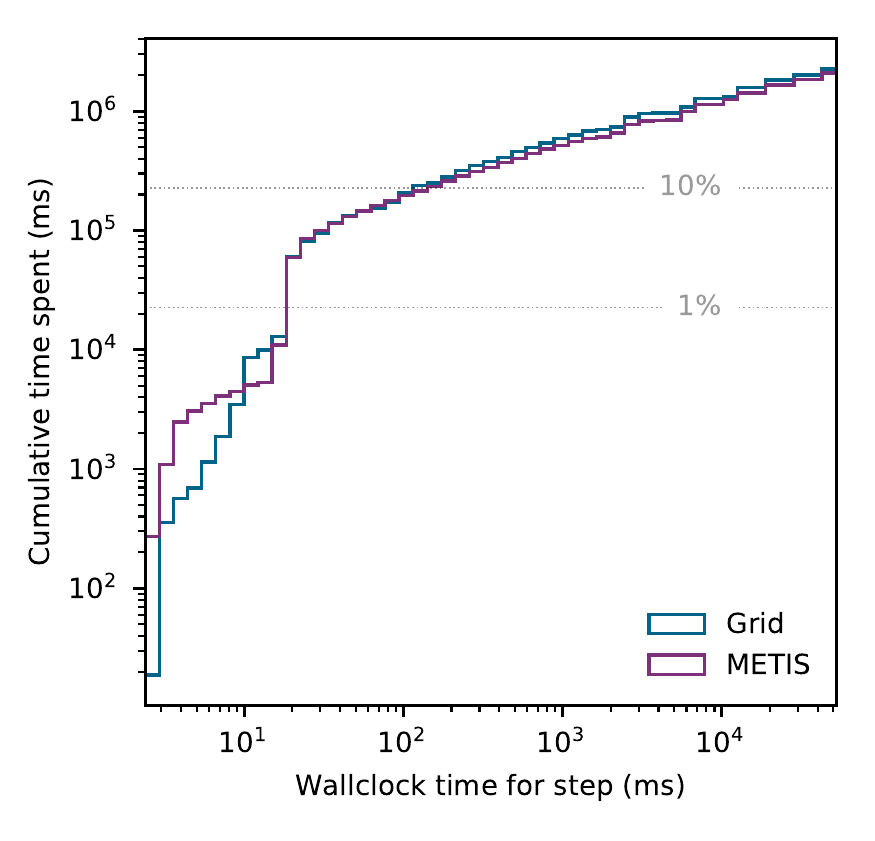}
	\caption{\textit{(left)} the wall-clock time per time-step as a function of the number of active particle for the EAGLE-25x8 box, i.e. a replication of the box used for Fig. \ref{fig:problem:strong} twice along each axis on 64 cores (4 nodes of the COSMA-5 system). This setup has $4\times10^8$ particles, and ensures that there is exactly the same initial particle load and distribution on each of the compute nodes (ideal weak-scaling test-problem). The deepness of the colour shows the number of steps in that bin; blue cells show where time-steps lie in the plane when using the `grid' strategy, and purple cells show the same but for the `METIS' domain decomposition strategy. The `main sequence' of steps at high particle counts scales very well and display the expected $\mathcal{O}(N)$ behaviour, but at lower active particle counts the fixed cost of each time-step is most apparent. The METIS domain decomposition strategy allows \swift{} to reduce all communications in these small steps to a minimum. Here, that is expressed through the movement of the cluster of points between 1 and 1000 active particles, which now require a much lower wall-clock time for computation. \textit{(right)} Cumulative histogram of the time spent in the different steps. Note that the short-running tasks (less than 10ms or so) contribute very little to the overall run time (less than 1\%) in a pure SPH problem, despite their high number. The grey box on the left panel encompasses the steps that are involved in rebuilding the space, cells, and tasks. These are not included in the histogram plot.\vspace{-0.2cm}}
    \label{fig:domain:tasks}
\end{figure*}

The above-mentioned `grid' and `METIS' strategies are compared in Fig. \ref{fig:domain:tasks}. The majority of the run time of a given SPH-only \swift{} run is spent in the long-running time-steps; ensuring that these are well load-balanced is still key. This also means that, in general, the overall run time of a given SPH-only simulation is impacted little by the addition of a more complex domain-decomposition strategy, and can actually have a negative impact on the time-to-solution if too much weight is given to the small time-steps.  Unfortunately, the effect of adding more sub-grid physics modules to \swift{}, such as cooling, will be to increase the dynamic range of time-steps and move more steps into the lower time-step bins. This makes reducing the cost of a lower number of particle updates essential in a code that targets galaxy-formation problems, where sub-grid physics is necessary. The `METIS' strategy manages to reduce the fixed cost of a small time-step by an order of magnitude by ensuring that those regions do as little MPI communication as possible, working mainly with particles stored locally.

%% file: results.tex
\begin{figure}
    \centering
    \includegraphics[width=\columnwidth, trim=0cm 0.5cm 0cm 0.5cm]{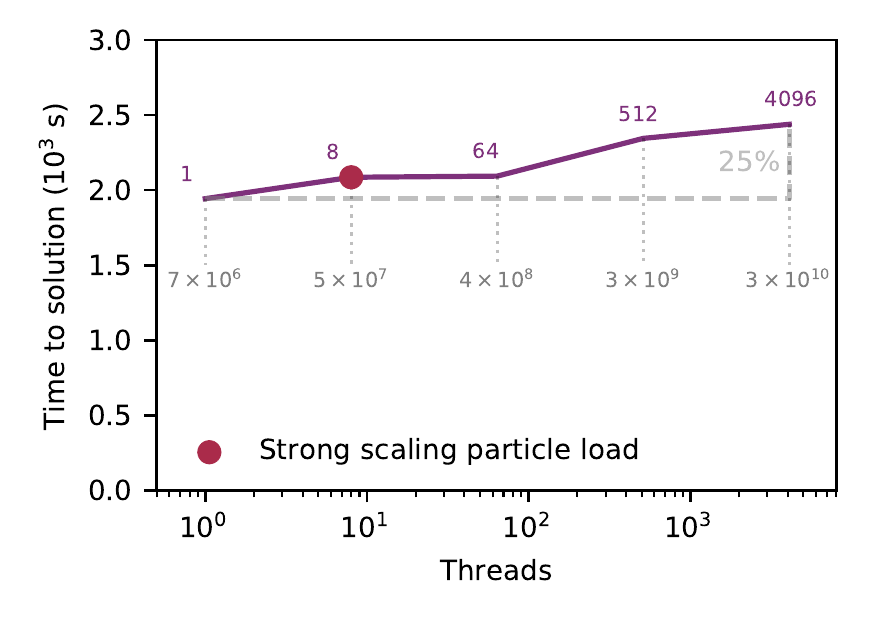} \caption{Weak-scaling results with \swift{} on a representative cosmological problem. Here, there is a particle load of $7\times10^6$ particles per core over $8197$ steps and 1 MPI rank \emph{per socket} (2 ranks per node). The particle data is taken from the EAGLE simulations \cite{Schaye2015} from a snapshot at redshift $z=0.1$ near the end of the calculation. For convenience, the particle loads are indicated for each core count in grey. This plot uses the METIS `costs/tasks' domain decomposition strategy, and as a result shows near-perfect weak scaling despite the large dynamic range and the limited available work per thread. These problems are highly adaptive, with a similar timestep hierarchy to the one shown in Fig. \ref{fig:problem:dynamic}. Note that the jump from $64$ to $512$ threads occurs when the calculation starts using nodes that are distant in the network topology. The red dot indicates the problem size used for the strong-scaling tests (Fig. \ref{fig:problem:strong} and \ref{fig:swift:strong}).}
    \label{fig:result:weak}
\end{figure}

The set of above optimizations, from the inclusion of multiple time-steps, a task-based domain decomposition scheme, and only drifting active particles thanks to the in-built adaptive mesh scheme, give \swift{} excellent weak-scaling characteristics in such an adaptive environment (see Fig. \ref{fig:result:weak}). 

All results were obtained on the COSMA-5 DiRAC2 Data Centric System, located at the University of Durham. The system consists of 420 nodes with 2 Intel Sandy Bridge-EP Xeon E5-2670 (8 physical cores with AVX capability) at 2.6 GHz with 128 GByte of RAM. The nodes are connected using Mellanox FDR10 Infiniband in a 2:1 blocking configuration. 

The effect of using METIS to decompose the domain is twofold. First, the particles on the smallest time-steps no longer need to perform MPI communications. Secondly, all particles are better decomposed as each node has a balanced amount of work to perform in these steps; this is because the \swift{} scheme partitions \emph{work}, rather than \emph{particles}. These are both ensured by using the `tasks/costs' strategy.

In terms of raw performance, on steps where all particles are active, \swift{} reaches an average of $5.8\times10^{-6}~\rm{s}$ per update on 1 core for the $53\times10^6$ particle case shown in Fig.~\ref{fig:swift:strong}. Although \swift{} uses a simple SPH implementation (i.e. without a Riemann solver), this compares favourably to the results of \cite{Oger2016}, who reported wall-clock times per update around $1.7\times10^{-4}~\rm{s}$. This demonstrates the effectiveness of the neighbour finding algorithm in \swift{} and the SIMD implementation which compensate for the over-heads introduced by the logic necessary to only drift and kick the relevant particles. On $32$ cores, the performance drops slightly to  $7.1\times10^{-6}~\rm{s}$ per update per core. Analysing all 4096 steps, including the ones that only update a handful of particles, on 32 cores \swift{} averages at $1.9\times10^{-5}~\rm{s}$ per update per core. Finally, on 4096 cores, with $25\times10^9$ particles (Fig.~\ref{fig:result:weak}), \swift{} achieves $2.4\times10^{-5}~\rm{s}$ per update per core. Higher performance is achieved on systems where AVX2 and AVX-512 instruction sets are available \cite{Willis2017}.

%% file: conclusion.tex
In this work, the problem of cosmological simulations as an example of extreme adaptivity has been explored, and multiple time-stepping has been shown to be a necessity when performing such a simulation. Without multiple time-stepping, cosmological simulations would present infeasible run times of over 100 times that of what a multiple time-step scheme is able to provide. Time-to-solution has been presented as an adequate replacement for `FLOPS' as a relevant performance metric for adaptive problems. The use of an adaptive cell grid has been shown to reduce significantly the computation required by allowing efficient neighbour searching using a pseudo-Verlet list. The cell structure also allows only the relevant particles to be drifted each time-step. Such an efficient scheme then presents issues for domain decomposition, and the use of the METIS graph decomposition library to partition the work has been shown to produce an improved decomposition that is more relevant to a code that uses asynchronous communications, leading to an improved time-to-solution.

\swift{} has been shown to provide both a significantly faster time-to-solution (over 30x in SPH-only mode) than the de-facto standard code in the cosmology/galaxy formation space, as well as weak-scaling almost perfectly up to 4096 codes on a highly adaptive cosmological problem. \swift{} is available for use by the community and is in open development.

%% file: acknowledgements.tex
The authors would like to thank the whole \swift{} team for their efforts, with particular recent contributions from Alexei Borissov, Aidan Chalk, Loic Hausammann, Bert Vandenbroucke, and James Willis. \swift{} is a multi-institution team effort and none of this work would have been possible without the efforts of our contributors. For this reason all authors have been listed alphabetically. 

JB is supported by STFC studentship ST/R504725/1. MS is supported by the NWO VENI grant 639.041.749. This work was supported by the Science and Technology Facilities Council (STFC) ST/P000541/1, and by {\sc intel} through the establishment of the ICC as an {\sc intel} parallel computing centre (IPCC). This work used the DiRAC Data Centric system at Durham University, operated by the Institute for Computational Cosmology on behalf of the STFC DiRAC HPC Facility (\url{www.dirac.ac.uk}). This equipment was funded by BIS National E-infrastructure capital grant ST/K00042X/1, STFC capital grant ST/H008519/1, and STFC DiRAC Operations grant ST/K003267/1 and Durham University. DiRAC is part of the National E-Infrastructure.